\documentclass{aa}
\AtBeginDocument{%
  \nolinenumbers
  \let\linenumbers\relax
}
\usepackage{graphicx} % Required for inserting images
\usepackage{natbib}
\bibpunct{ (}{) }{;}{a}{}{,}
\usepackage{txfonts}
\usepackage{lipsum}
\usepackage{amsmath}
\usepackage{siunitx}

\DeclareSIUnit \parsec {pc}
\DeclareSIUnit \erg {erg}
\DeclareSIUnit{\photon}{ph}

\begin{document}

\title{Insights on the Gamma-Ray Bursts variability in their cosmological rest frame}

\author{G. Della Casa\inst{1}\thanks{Corresponding author: giovanni.dellacasa@inaf.it} \and F. Fiore\inst{2} \and G. Dilillo \inst{3,4} \and S. Puccetti\inst{5} \and A. Vacchi\inst{6,7}}

\date{Received date /
Accepted date }

\institute{Istituto Nazionale di Astrofisica - Istituto di Astrofisica e Planetologia Spaziali, Via del Fosso del Cavaliere 100, 00133 Roma, Italy \and Istituto Nazionale di Astrofisica - Osservatorio Astronomico di Trieste, Via G.B. Tiepolo 11, 34143 Trieste, Italy \and Istituto Nazionale di Astrofisica - Osservatorio Astronomico di Roma, Via Frascati 33, 00078 Roma, Italy \and Agenzia Spaziale Italiana - Space Science Data Center, Via del Politecnico, 00133 Roma, Italy \and Agenzia Spaziale Italiana,  Via del Politecnico sec, 00133 Roma, Italy \and Università degli Studi di Udine, Via A. Palladio 8, 33100 Udine, Italy \and Istituto Nazionale di Fisica Nucleare - sez. Trieste, Località Padriciano 99, 34149 Trieste, Italy}

\abstract{Gamma-ray bursts temporal profile can be extremely variable, going from a single pulse of a few seconds duration to multiple superimposed pulses occurring over tens or even hundreds of seconds. The variability displayed in the lightcurve of each gamma-ray burst can be the result of the activity taking place in the central engine that generates these violent phenomena, as well as due to magnetic reconnection activities at larger distances. The objective of this work is to find the shortest variability hidden in the lightcurves of the GRBs, with particular focus for the ones with measured redshift, on timescales as short as few milliseconds. This variability will then be related to physical characteristics of the central engine, and evidences of its relation with the spectral parameters of the burst, such as the isotropic energy and peak energy, will be presented. This research is even more relevant in view of the future generation of satellites with improved timing resolution, that will allow us to explore the possible variability in the microsecond region.}


\maketitle

\section{Introduction}
\label{sec:intro}

Gamma-ray bursts (GRBs) are the most luminous events (more than $10^{53}$ ($\Omega/4\pi$) erg released in an isotropic emission model) in the Universe. GRBs are thought to be produced by the collapse of massive stars and/or by the coalescence of two compact objects. Observed for the first time nearly 60 years ago \citep{1973ApJ...182L..85K}, these explosions are still the focus of numerous missions to improve our understanding of the physical processes that generate them. The other main feature of GRBs, besides their enormous brightness, is their exceptional variability, often as short as few ms \citep{2013MNRAS.432..857M, 2014ApJ...787...90G}. These characteristics led to the development of the fireball model, i.e. a relativistic bulk flow, i.e. a powerful jet, where shocks efficiently accelerate particles \citep{1993ApJ...405..278M, 1999PhR...314..575P}. The energy dissipation of the ultra-relativistic particles in internal shocks then produces the observed prompt X-ray and gamma-ray emission. GRB prompt spectra can be reproduced reasonably well by synchrotron radiation in the marginally-fast cooling regime \citep{2017ApJ...846..137O, 2019A&A...628A..59O}, although, to make up for the difficulties in explaining the slope observed in the lower energy part of the spectra of GRBs compared to the synchrotron limit, photospheric models might come into the picture \citep{2019Galax...7...33P}. Further interactions of the jet with the Interstellar Medium then produce the so-called afterglow emission, discovered by BeppoSAX in X-rays \citep{1997Natur.387..783C}, and immediately seen from radio to gamma-ray energies \citep{1997Natur.386..686V, 1997A&A...327L..36P, 1998ApJ...497..288W}. The afterglow emission is reasonably well described by synchrotron radiation from the external forward-shock model \citep{1997ApJ...476..232M, 2001ApJ...560L..49P}.

Although broadly successful in explaining GRB observations, the fireball model implies a thick photosphere, hampering direct observations of the hidden inner engine that accelerate the bulk flow. We are then left in the frustrating situations where we see at work daily the most powerful accelerators in the Universe, but we are kept in the dark over their operation. One possibility of shedding light on their inner engines is through GRB fast variability.  Early numerical simulations \citep{1997ApJ...490...92K, 1998MNRAS.296..275D}, as well as modern hydro-dynamical simulations \citep{2010ApJ...723..267M}, and analytic studies \citep{2002ApJ...572L.139N, 2002MNRAS.330..920N, 2002MNRAS.331...40N} suggest that the GRB light-curves reproduce the activity of their inner engines.

One of the most important open questions in GRBs is the composition of the jets. It is still not clear, indeed, whether the jet energy is mostly carried out from the central engine in the form of Poynting flux (magnetic jet) or as kinetic energy of the matter (baryonic jet). In the two scenarios, the mechanism for extraction of the jet energy is very different, being dominated by magnetic reconnection events in the first case, and by internal shocks in the second case. The two scenarios predict different variability properties. Fast variability should be related to the central engine activity in the matter dominated model, while in the magnetic model it reflects the possible presence of turbulent regions with fast proper motions within the same emission region. 

Gamma-ray burst light-curves have been investigated in detail down to hundreds of microseconds \citep{2013MNRAS.432..857M, 2015ApJ...811...93G, 2023A&A...671A.112C, 2025A&A...702A..95M}, searching for the shortest timescale over which we observe a significant variation of the GRB profile, from here on also called minimum variability timescale (MVT), $\Delta t_{min}$. This is an important indicator of the sharpness of the lightcurve pulses, of the activity and dimension of the GRB central engine and could be used to constrain theoretical models proposed to explain the prompt phase.

In this paper, we study the MVT of thousands of GRBs observed by Fermi-GBM across a broad energy band. Building on established Haar wavelet methods (e.g. \citet{2013MNRAS.432..857M}, \citet{2014ApJ...787...90G}) we extend the analysis to a larger sample, focusing in particular on the 150 GRBs of this sample with measured redshifts. This allows us to examine MVT in both the observer frame and the cosmological rest frame.

The article is structured as follows: in Sect.~\ref{sec:methods} we present the theoretical background of the technique used to retrieve the MVT, then in Sect.~\ref{sec:sample_sel} we describe the two datasets that are employed for this study and, in Sect.~\ref{sec:data_prep}, the data processing pipeline; subsequently, in Sect.~\ref{sec:results} we discuss the results obtained, both in the observer and cosmological rest frames, and possible correlations to other characteristic spectral values of GRBs; finally, we conclude by summarizing the principal findings of this work.

\section{Methods}
\label{sec:methods}

A possible technique used since the beginning of this century to extract the shortest variability of a GRB is the wavelet analysis, in particular the Haar discrete wavelet transform \citep{2000ApJ...537..264W, 2013MNRAS.432..857M, 2014ApJ...787...90G}. Using a wavelet transform allows you to obtain information on the signal variation in both time and frequency contrary to a Fourier transform that is limited by the Fourier uncertainty principle. There are two types of wavelet transform: continuous and discrete. For the purpose of this work, the discrete wavelet transform is the obvious choice, working with a discrete set of data. A discrete wavelet is a set of functions that forms an orthonormal basis, and can be used to explore the time-structure of the GRB spectrum. From the coefficients generated by the transform, it is very simple to separate the noise (low value) from the signal (high value) by setting a threshold and preserving only the signal coefficients \citep{1997ApJ...483..340K, 2018TEMA.Brassarote}. The simplest discrete wavelet existing is the Haar wavelet \citep{1910MATH.Haar}, which consists in an anti-symmetric function, setting one bin positive, the following one negative and all the others equal to zero. An example is given supposing to have a scale of $\frac{1}{2}$:

\[{\psi(t)}=
\begin{cases}
    +1 \text{ if } 0 \leqslant t < \frac{1}{2}\\
    -1 \text{ if } \frac{1}{2} \leqslant t < 1\\
    0 \text{ everywhere else}
\end{cases}\]

The discrete wavelet transform works for a set of data $\{ X_1; X_2; ...; X_N \}$ of dimension N that must be a power of 2, i.e. $N = 2^q$ ($q \in \mathbb{Z}^{+}$). The Haar transformation returns the same amount $N$ of coefficients, comprising $N - 1$ detail coefficients, obtained from the difference between adjacent bins, for an increasing scale $k = 2^i$, with $i$ in $\{0; 1; 2; ...; q-1\}$ (shown in equation \ref{detail_coef}), while the last coefficient is the mean value of the set of data.

\begin{equation} \label{detail_coef}
d_{j,k} = \frac{1}{\sqrt{2k}} \left[\sum_{n=0}^{k-1}X_{2jk-n} - \sum_{n=0}^{k-1}X_{2jk-k-n}\right]
\end{equation}

where j is an index that can take the following values $\{1; 2; 3; ...;N/2k\}$.

As accurately demonstrated in \citet{2007ApJ...667.1024K} and \citet{2014ApJ...787...90G}, it is possible to associate the Haar wavelet coefficients to the structure function, $SF$, defined as:
\begin{equation} \label{struc_func}
SF(\tau) = \langle \left[ X (t) - X (t+\tau) \right] ^2 \rangle
\end{equation}

where the brackets indicates the mean value over the full data sample, $\tau$ is the binning time and $X$ is the count rate for each bin. To make the structure function analysis more robust, the Haar wavelet transform is not applied only once but N times, each time on the same set of data (of dimension N) circularly shifted by one position. The final value of $SF$ is obtained by averaging the N transforms results, shifted back in their original position. This is called a non-decimated Haar wavelet transform, which is translation invariant and solves artifacts connected to this type of wavelet transform.

\begin{figure*} [htb!]
\centering
\includegraphics[width=\columnwidth]{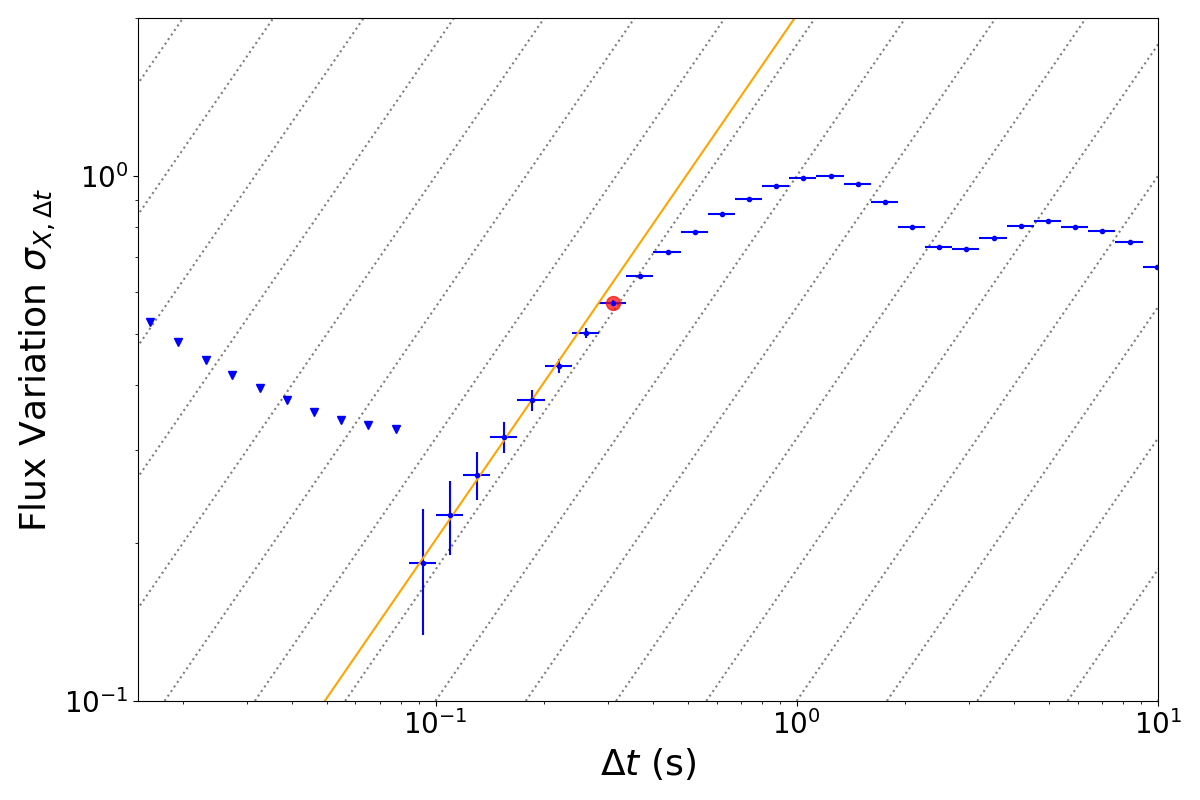}
\includegraphics[width=\columnwidth]{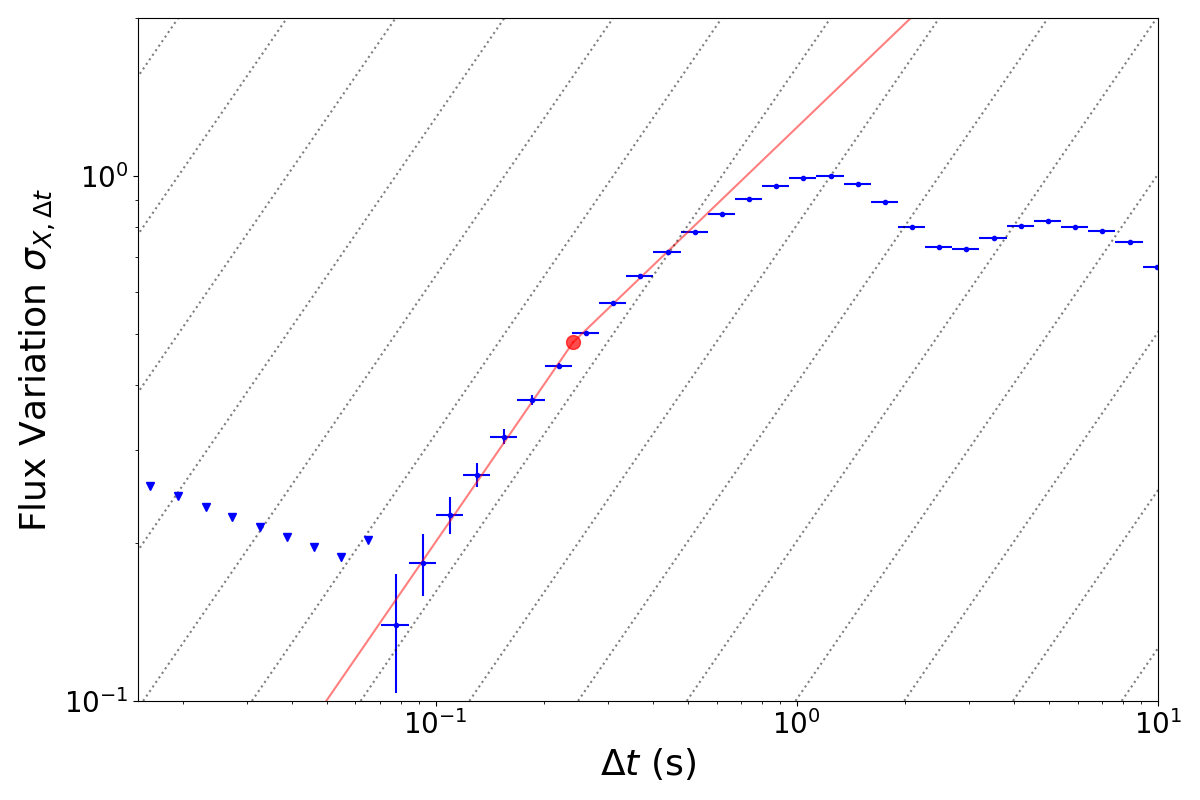}
\caption 
{ \label{MVT} 
Example of MVT calculation for GRB080730B, using the two methods. \textit{Left image}: method 1, the orange line corresponds to the best fit of the linear phase. The red dot shows the timescale corresponding to the end of the linear phase, i.e. the MVT. \textit{Right image}: method 2, The red line shows the initial linear phase, followed by a flatter phase, with the intersection highlighted by the red dot. The triangles represent the $3\sigma$ Poisson noise threshold for the timescales where $\sigma_{X,\Delta t}$ is lower than this limit. In the case presented here the results are very similar, but they can differ as illustrated in Sect.~\ref{subsec:obs_frame}. The black dotted lines help the eye to identify the power-law scaling.}
\end{figure*} 

The main difference for the computation of the MVT in the previously cited articles, is the criterion used to determine, from the scaleogram, the shortest timescale of significant variation in the GRB lightcurve. While \citet{2000ApJ...537..264W} and \citet{2013MNRAS.432..857M} consider the shortest variability the first timescale for which the structure function is above the noise level, \citet{2014ApJ...787...90G} defines the MVT as the first timescale of uncorrelated temporal variability, i.e. the first timescale for which the square root of the structure function, $\sigma_{X,\Delta t}$ is not growing linearly. Since the observed noise changes from one detector to the other, depending for example on the collecting area, a method that relies on the noise (as it is for the first two studies cited above) would give results that relies on the instrument itself. Instead, the criterion employed in the last cited work is independent from the noise measured by the instrument, thus providing the most suitable result.

This work considers two possible methods to find the MVT, which are shown in Fig.~\ref{MVT}. The first method (henceforth referred to as method 1, shown in the left figure) involves finding the best linear fit for the linear rising phase of $\sigma_{X,\Delta t}$, according to a $\chi^2$ test ($\alpha = 0.05$). As more timescales are considered, the fits $\chi^2$ initially improve (while in the linear growth phase), but then deteriorate as the curve starts to flatten. The second method is taken from \citet{2015ApJ...811...93G}, and was provided via private communications with Dr. N. R. Butler (called method 2). Looking at the scaleogram (right-hand figure), the MVT is given by the intersection between two lines, one fitting the linear phase, the other one the subsequent flatter region.
 
\section{Sample selection}
\label{sec:sample_sel}  

For the study presented here, we used one of the largest catalogs of GRBs available, the Fermi Gamma-ray Burst Monitor (GBM) one. Since its launch in 2008, GBM was able to observe more than 3000 bursts in the $\SI{8}{\kilo\electronvolt}$--$\SI{40}{\mega\electronvolt}$ energy range, with a time resolution of $\SI{2}{\micro\second}$. 
This study was conducted using the GRBs detected by Fermi GBM between the 2008 July $14^{th}$ and 2021 October $18^{th}$. During this period of time, the detectors on board the Fermi satellite observed 3155 bursts. To ensure consistent results consistent for each GRB, if the time triggered event (TTE, described in the next section) data files of the detectors do not contain the full GRB or enough data to estimate properly the background, that GRB is discarded. This is the case for 93 GRBs, leaving us with 3062 GRBs, divided in this work in 511 short and 2551 long according to their duration, $T_{90}$ (the time interval over which $90\%$ of the photons are detected). The whole set of GRBs is initially presented in the following paragraph (Sect.~\ref{subsec:obs_frame}), but our real objective is to examine the GRBs in their cosmological rest frame. This is only possible for those with a well-established redshift.
The ``redshifted" set, that is used in Sect.~\ref{subsec:cosm_frame} and \ref{subsec:variance}, contains 150 GRBs, divided into 14 short and 136 long. Depending on the feature investigated, this sample may have been adapted in some subsections. The redshifts were mainly measured spectroscopically (only six of them are photometric redshifts). Note that the long/short division has been done keeping the $T_{90}$ of the observer frame, as explained in Sect.~\ref{subsec:cosm_frame}.

\section{Data processing}
\label{sec:data_prep}  

\begin{figure} [ht]
   \begin{center}
   \includegraphics[width=\linewidth]{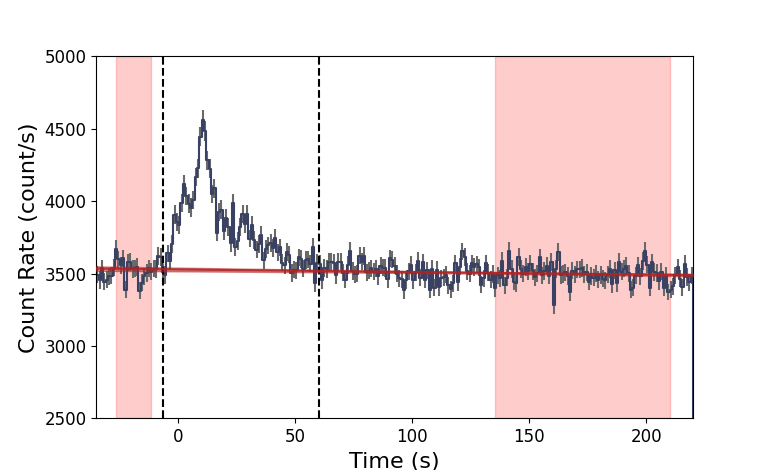}
   \end{center}
   \caption 
   { \label{GRB_bkg} 
    Example of the procedure of background estimation for GRB200402. The red line represents the linear fit of the background measured in the two shadowed regions. The dashed black lines represent the start and end of the GRB.}
\end{figure}

The first step consists in recovering the TTE data files of the bursts. The TTE data contains the temporally unbinned data, i.e. it provides information on the time at which every photon is detected, among other things. The GBM data tools proved to be very helpful for this task, providing a catalog with all the information ($T_{90}$, start time, fluence, ...) usually required to analyze a GRB and also a direct access to the archive where the TTE files are stored to promptly download them, all through a simple Python API. For each triggered event, there is one file for each detector on board of Fermi GBM (this means 12 files for the NaI detectors, covering the $\SI{8}{\kilo\electronvolt}$--$\SI{1}{\mega\electronvolt}$ energy range, and 2 for the BGO ones, which are not used for this work). This catalog can also be used to obtain a list of the triggered detectors for each GRB. Data from untriggered detectors are filtered out, while data from triggered detectors are merged into a single file. By fixing the energy band and the binning timescale we obtain the GRB lightcurve from this merged file. Except for the analysis reported in Sect.~\ref{subsec:obs_frame}, we always use the full energy range covered by the NaI detectors, 8--$\SI{1000}{\kilo\electronvolt}$. For the binning timescale, we have chosen to start from $\SI{0.2}{\milli\second}$. 
The second step that we perform before analyzing the lightcurve is the removal of the background. For this purpose we use the \textit{gbm.background} routine from the GBM tool package. It first requires two time intervals, one before and one after the burst, in order to provide the best fit of the background during the GRB prompt emission phase. These two sets of data are taken directly from the same merged file produced for the GRB lightcurve, one going from 20 to 5 seconds before the burst and one from 75 to 150 seconds after the burst. For the analysis conducted here, a linear fit perfectly suits the requirement. An example is shown in Fig.~\ref{GRB_bkg}. This fit is then subtracted from the burst itself. 

\begin{figure*} [hbt]
\centering
\includegraphics[width=\columnwidth]{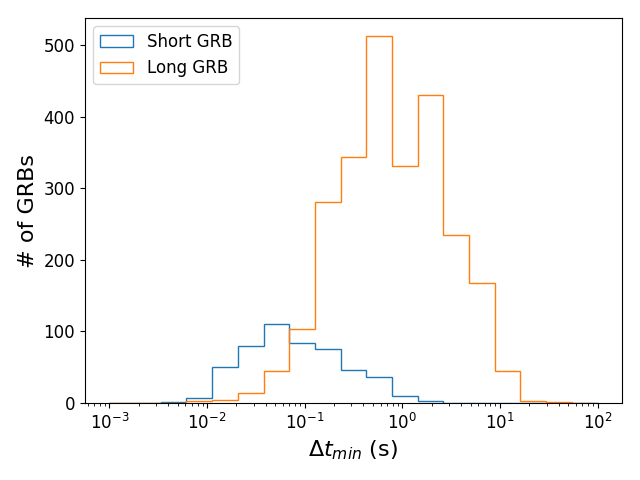}
\includegraphics[width=\columnwidth]{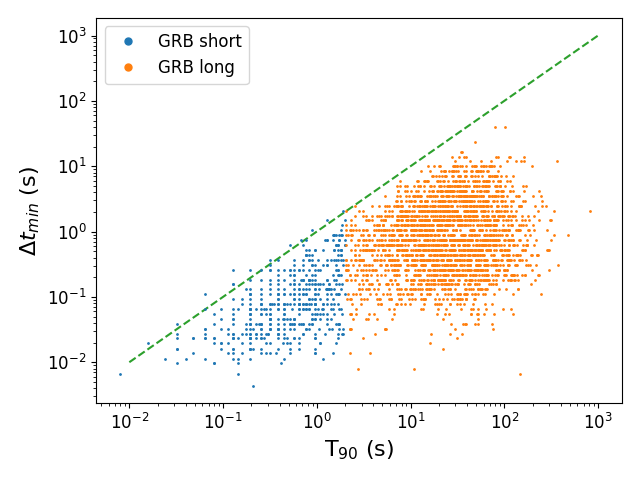}
\caption 
{ \label{T90_MVT} 
MVT computed with Method 1. \textit{Left image}: distribution of the MVT for both long and short GRBs. \textit{Right image}: MVT versus the T90 for long and short GRBs. The dashed green line represents the $T_{90} = \Delta t_{min}$ relation. The errorbars on both x and y axes of the few points above this line intersect with this line justifying the fact that the $\Delta t_{min}$ is apparently longer than the burst itself.}
\end{figure*} 

We then apply the methods described in Sect.~\ref{sec:methods} to the logarithm of this lightcurve, in order to obtain the corresponding scaleogram, and finally identify the shortest uncorrelated time variability using the methods described in Sect.~\ref{sec:methods}. 

\section{Results}
\label{sec:results}  

\subsection{MVT in observer frame}
\label{subsec:obs_frame}

The median value found for long GRBs using Method 1 is $\SI{740}{\milli\second}$, with values ranging from $\SI{6.6}{\milli\second}$ to $\SI{40}{\second}$, and for short ones is $\SI{78}{\milli\second}$, with values contained between $\SI{4.4}{\milli\second}$ and $\SI{2.1}{\second}$. Instead with Method 2, the median value are $\SI{583}{\milli\second}$ for long GRBs, with values ranging from $\SI{6.2}{\milli\second}$ to $\SI{130}{\second}$, and $\SI{53}{\milli\second}$ for short GRBs, with values contained between $\SI{2.3}{\milli\second}$ and $\SI{2.4}{\second}$. The values obtained with the Method 2 are higher than those presented in \citet{2015ApJ...811...93G} with a nearly identical method because in our work the upper limit values are considered as the actual MVT of the bursts in the calculation of the median, contrary to what was done in the cited work, where the median was calculated using a Kaplan-Meier survival analysis.

\begin{figure} [htb!]
\centering
\includegraphics[width=\linewidth]{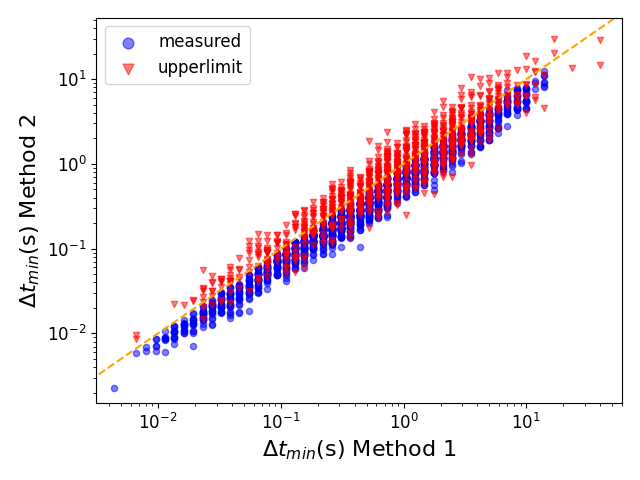}
\caption 
{ \label{T90_MVT_B} 
Comparison of the MVT values obtained for each GRB, both short and long ones, with the two methods. When only an upper limit is available for the Method 2, a red triangle is plotted. The yellow dashed line represent the identity function.}
\end{figure} 

Considering only the measured MVTs, the median values are $\SI{410}{\milli\second}$ and $\SI{41}{\milli\second}$ for long and short GRBs, respectively. The distribution of the values is shown in Fig.~\ref{T90_MVT} and Fig.~\ref{T90_MVT_B}. These results show that, despite the similar distribution shapes, method 2 yields shorter MVT values (usually a factor 1--3). However, it also has several GRBs for which it can only provide an upper limit value for the MVT. Therefore, we can assert that method 1 is more conservative. Furthermore Fig.~\ref{T90_MVT} displays for method 1 the MVT of each burst and its duration, $T_{90}$. No information on any relation between these quantities can be extrapolated from here, but it is useful for demonstrating that the MVT calculated with method 1 is not larger than the duration of the GRB itself, which would make no sense.

\subsection{MVT in cosmological rest frame}
\label{subsec:cosm_frame}
The results shown in the following sections will display only the MVT obtained with method 1 since we have already explained that the main difference between the two methods consists simply in a shift towards lower values of MVT for method 2, but the distribution is similar, and the upperlimit cases are removed. 

As we are now operating in the cosmological rest frame, the $\Delta t_{min}$ that was calculated in the observer frame should be re-evaluated accounting for time dilation. However, a possible bias could be introduced by the choice of adopting a simple time dilation correction, without taking into account some other effects, such as the narrowing of the peaks for GRBs located at higher redshift, since they are observed in a harder energy band \citep{1995ApJ...448L.101F}. These effects could cause a milder dependence of the time dilation from the redshift. In particular, if we assume the relation between the narrowing of the pulse and the energy band considered, quantified by \citet{1995ApJ...448L.101F} as $\Delta t \propto E^{-0.4}$, as a possible correction, then the time dilation would be accounted by dividing the MVT by $(1+z)^{0.6}$. We decide to use this correction to obtain the MVT in the cosmological rest-frame, defined as $\Delta t_{min,z}$. Concerning the $T_{90}$, we do not apply any correction moving from the observe frame to the cosmological rest frame since we must account for the "tip-of-the-iceberg" effect \citep{2013ApJ...765..116K}, i.e. the reduction of the SNR for increasing redshifts that highly impacts the measured duration of the GRB. This effect is less relevant for the $\Delta t$, so we decided to not account for it, but we should not forget about it in the following steps. For this new set of MVTs, the median value for long GRBs is $\SI{300}{\milli\second}$, with a minimum value of $\SI{4.8}{\milli\second}$ and a maximum of $\SI{12}{\second}$, while for short GRBs it has become $\SI{35}{\milli\second}$, with a minimum value of $\SI{8.9}{\milli\second}$ and a maximum of $\SI{367}{\milli\second}$.

\subsubsection{Narrowness-hardness relation}

Over the last few decades many studies have been pointing out that, when examining the spectrum of a GRB, the peaks tend to become narrower for higher energy bands (e.g. \citet{1995ApJ...448L.101F, 1996ApJ...459..393N}). Consequently, it is expected that the same happens for the MVT, since it was demonstrated before that the value found for the MVT corresponds to the shortest rise time fluctuation. Since we now have a significant sample of GRBs with redshift, we perform this study in the cosmological rest frame. The full energy range is divided in two bands (in the cosmological rest frame): the first one goes from $\SI{20}{\kilo\electronvolt}$ to $\SI{100}{\kilo\electronvolt}$ (named soft energy band), the second one from $\SI{100}{\kilo\electronvolt}$ to $\SI{1000}{\kilo\electronvolt}$ (hard energy band). We ignore the first part of the full energy band (8--$\SI{20}{\kilo\electronvolt}$) since the effective area of the detectors drops abruptly at these energies.

\begin{figure} [ht!]
\centering
\includegraphics[width=\linewidth]{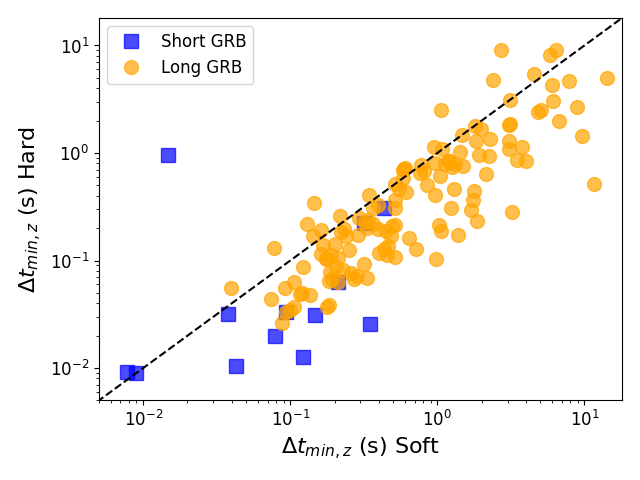}
\caption 
{ \label{Micr_energy} 
Comparison of the values of the MVT for a same GRB using different energy bands. The ``Soft" $\Delta t_{min,z}$ was calculated between $\SI{20}{\kilo\electronvolt}$ -- $\SI{100}{\kilo\electronvolt}$, while the ``Hard" $\Delta t_{min,z}$ was calculated between $\SI{100}{\kilo\electronvolt}$ -- $\SI{1000}{\kilo\electronvolt}$, for both long, orange circles, and shorts, blue squares, GRBs. Unless otherwise stated, this legend is valid for all the following plots. The black dashed line represents the identity function.}
\end{figure} 

In Fig.~\ref{Micr_energy}, we observe a clear majority of cases where the MVT is shortened in the hard energy range as compared to the soft energy range. This confirms the sharpening of at least the narrowest peak, but likely also of all the GRB temporal profile, and further justifies our choice of adopting the "Fenimore-like" correction when accounting for the time dilation. Considering the few GRBs that do not follow this trend, the majority of them have MVT values very similar in both cases while only a single case is completely off.
However, there is a possible explanation to this exception. As demonstrated in \citet{2014ApJ...787...90G}, faint GRBs are expected to have a longer MVT due to the difficulties of measuring it. The GRB in question (GRB 160821) is extremely faint at high energies, while clearly visible in the low energy band. Consequently we can assert that the sharpness-hardness relation is true provided that we are able to properly estimate the MVT at all energies, i.e. the GRB is not faint in any energy band.

\subsubsection{Relation between the redshift and MVT}

One of the first aspect that can be investigated, similarly to what was done in \citet{2014ApJ...787...90G} and \citet{2015ApJ...811...93G}, is the eventual relation between this MVT and the redshift of the GRB, as reported in Fig.~\ref{z+1_microvar}, to confirm the goodness of the relation used to recalculate the MVT in the cosmological rest frame.

\begin{figure} [ht!]
   \centering
   \includegraphics[width=\linewidth]{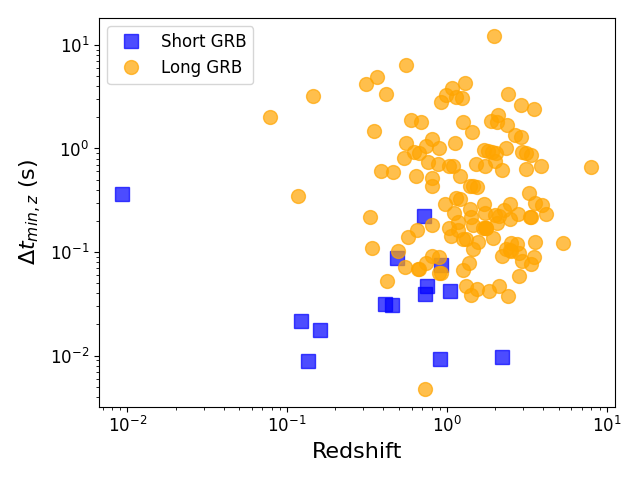}
   \caption 
   { \label{z_microvar} 
    Plot showing the redshift vs $\Delta t_{min,z}$ comparison for both long and short GRBs.}
\end{figure}

\begin{figure} [ht!]
   \centering
   \includegraphics[width=\linewidth]{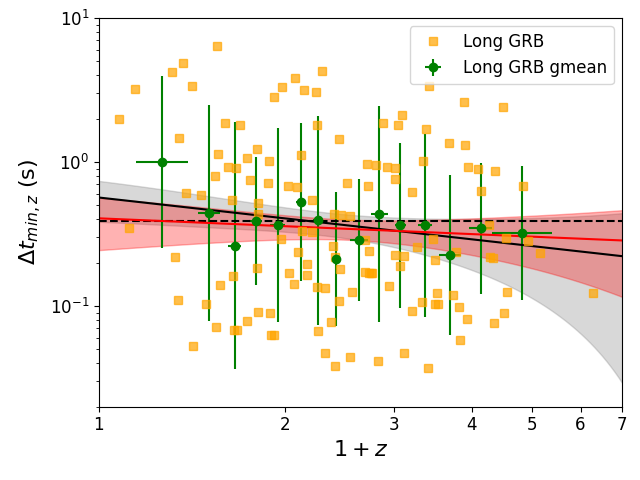}
   \caption 
   { \label{z+1_microvar} 
    Study of the time dilation through the correlation between $1+z$ and the redshifted MVT for long GRBs, binning them and using the geometrical mean. The black solid line represents the best linear fit, while the dashed black line corresponds to the $\Delta t_{min} \sim (1+z)^{0.6}$ relation. The grey shadowed region individuates the $2\sigma$ confidence region around the fit. The red solid line and corresponding shadowed region refer to a fit performed without the first bin of GRBs.}
\end{figure}

At first glance, if we look at Fig.~\ref{z_microvar} there does not seem to be any relation between the redshift and the shortest variability of the corresponding bursts, neither for long or short GRBs. However, more interesting features can be observed by binning together GRBs with similar redshift, as done in \citet{2014ApJ...787...90G} and \citet{2015ApJ...811...93G}. For the case presented here (Fig.~\ref{z+1_microvar}), we took the long GRBs and binned them together by groups of 9 (10 for the last group), using a geometrical mean. Performing a power-law fit the trend indicates that the $\Delta t_{min,z}$ decreases with the redshift, indicating that other effects might come into play, such as the mentioned "tip-of-the-iceberg" effect, when correcting the MVT for the time dilation. However, if we include the confidence interval up to $2\sigma$ (grey area in Fig.~\ref{z+1_microvar}), our fit is compatible with a non-dependence relation between the two quantities, in other words it is compatible with the milder cosmological time dilation that accounts for "Fenimore-like" correction ($\Delta t_{min} \sim (1+z)^{0.6}$). The situation is actually worsened by the very low redshift region, where the statistics is low. If we exclude it provisionally, the $2\sigma$ confidence region clearly confirms this result.

\subsubsection{Shortest variability vs isotropic energy}
\label{subsubsec:Eiso}

As already displayed in Sect.~\ref{subsec:obs_frame}, energy and sharpness of the lightcurve seem to be related. To explore the depth of this relation, we are interested in comparing the isotropic energy (i.e. the total energy emitted by the GRB if the emission is over 4$\pi$ steradians) of the GRBs, $E_{iso}$, with their MVT in the cosmological rest frame.

\begin{equation} \label{iso_energy}
E_{iso} = \frac{4\pi D_L^2 S_{bol}}{1 + z}
\end{equation}

where $D_L$ represents the luminosity distance of the GRB from the observer, calculated from the redshift using $H_0$ = $\SI{67.4}{\kilo\meter \per\second \per\mega\parsec}$, $\Omega_m = 0.315$ and $\Omega_{\Lambda} = 0.685$ (e.g. \citet{2020A&A...641A...6P}), and $S_{bol}$ is the bolometric fluence calculated between $\SI{10}{\kilo\electronvolt}$ and $\SI{1}{\mega\electronvolt}$, provided by the Fermi GBM catalog.

\begin{figure} [ht!]
   \begin{center}
   \includegraphics[width=\linewidth]{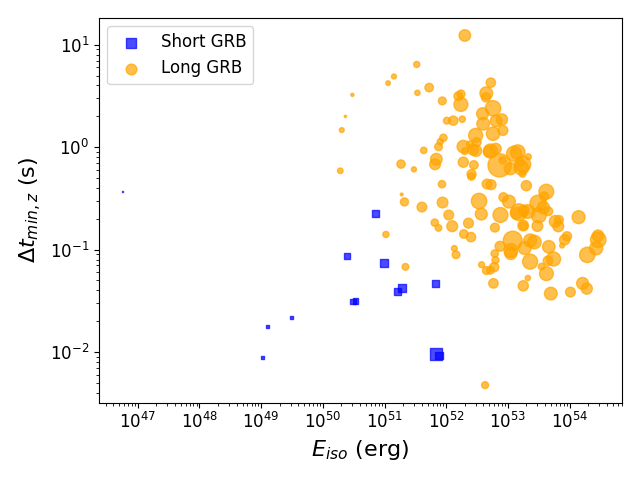}
   \end{center}
   \caption 
   { \label{Eiso_microvar} 
    Display of the dependence on the isotropic energy on the MVT. The dimension of the circles is proportional to the redshift of the GRB.}
\end{figure}

\begin{figure} [ht!]
   \begin{center}
   \includegraphics[width=\linewidth]{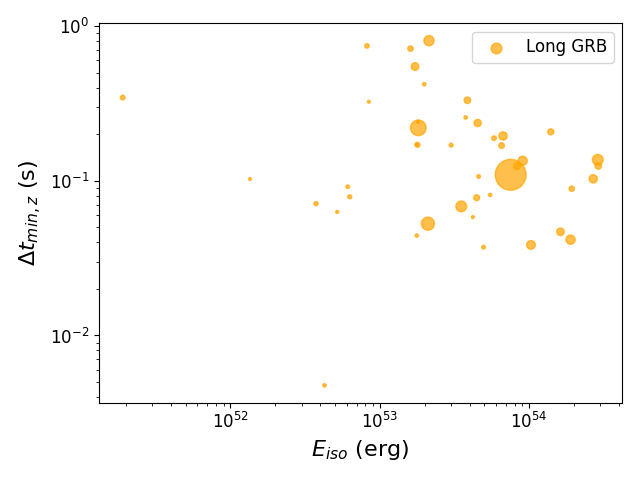}
   \end{center}
   \caption 
   { \label{Eiso_microvar_highflu} 
    Equivalent plot to Fig.~\ref{Eiso_microvar}, but taking into account only long GRBs with high fluence ($> \SI{2e-5}{\erg \per\square\centi\meter}$). The dimension of the circles is proportional to the fluence.}
\end{figure}

There are two main information to retain from the $E_{iso}$ vs $\Delta t_{min,z}$ plot (Fig.~\ref{Eiso_microvar}). The first one concerns long GRBs, where it is clear that higher isotropic energies correspond to lower MVT values, apparently following a power law trend. This is further confirmed calculating the Pearson correlation coefficient between the logarithm of these quantities, which returns $r = -0.51$ (with a p-value $< 10^{-3}$), indicating a probable anti-correlation. It could be asserted that this relation also holds true for the fluence. To this end, another study of $E_{iso}$ vs $\Delta t_{min,z}$ is provided, showing only the GRBs with high fluence ($> \SI{2e-5}{\erg \per\square\centi\meter}$). As it can be seen in Fig.~\ref{Eiso_microvar_highflu}, these GRBs are not all concentrated at high isotropic energies/low MVT, but the power-law behavior is still present, suggesting that the dependence of $E_{iso}$ on $\Delta t_{min,z}$ is genuine. It is important to highlight that GRBs with similar redshift display MVT values over 2-3 orders of magnitude, and similarly for the isotropic energy. This indicates that we are not limited by the Malmquist bias, that predicts that at higher redshifts only GRBs above a certain threshold of luminosity can be detected.

There are a few outliers, corresponding to GRBs with low redshift that display a lower isotropic energy ($< \SI{e51}{\erg}$) than what could be expected. It is speculated that these GRBs might be GRBs classified as long, but actually with a central engine characteristic of short GRBs.

The second interesting aspect that can be noticed is the evident separation between short and long GRBs. This can be actually expected, since short GRBs usually emit a lower total amount of energy, and we have already reported that the MVTs are shorter for short GRBs.

As it was demonstrated in \citet{2023A&A...671A.112C} and \citet{2025A&A...702A..95M}, the MVT that can be observed in a GRB depends on the peak rate of the narrowest peak. To verify if this relation can have an impact on the results found in this section, we performed a study in line with what was done in the two papers mentioned, described more in detail in Appendix~\ref{sec:appendixA}. This study allowed us to generate $10^4$ sets of GRBs with parameters similar to the real set but with $\Delta t_{min,z}$ and $E_{iso}$ completely unrelated. Using again the Pearson correlation test, we measured the degree of correlation between the two quantities for every set. We found that no set showed a comparable degree of correlation as the one found for the real set of data (the closest correlation found is -0.37).

\subsubsection{Shortest variability vs peak energy}

Another characteristic value of the GRB is the peak energy, $E_p$, the energy at which their $\nu$F$_{\nu}$ spectrum shows the highest value. In order to have a peak energy associated to it, the spectrum must be of one the two following types that are used in the Fermi GBM catalog: Band or Comptonized. The first model was introduced by \citet{1993ApJ...413..281B}, the second one consists in an exponentially attenuated power law. Unfortunately, requesting only the GRBs that have their spectrum best fitted with one of these two models further reduces the set of GRB with redshift available, since there are GRB spectra best fitted with only one power-law. Consequently there are only 90 GRBs, 6 short and 84 long, to analyze. The energy peak is calculated in the cosmological rest frame, $E_{p,z}$ by simply multiplying the value found in the catalog by $1 + z$.

\begin{figure} [ht!]
\centering
\includegraphics[width=\linewidth]{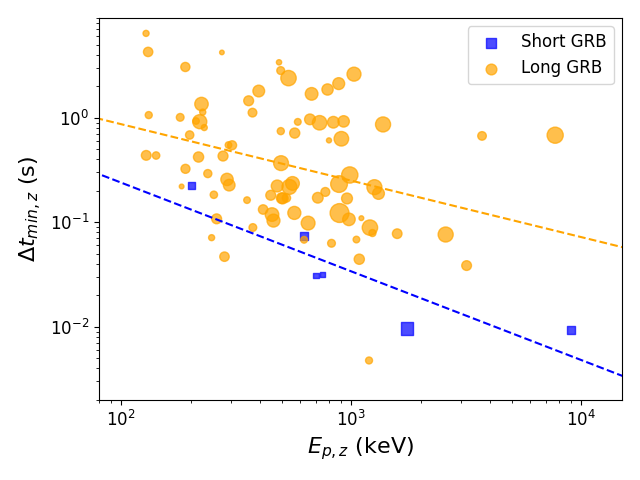}
\caption 
{ \label{Epeak_microvar} 
 Display of the relation between the peak energy and the MVT. The dimension of the markers is proportional to the redshift of the GRB.}
\end{figure} 

What is extrapolated from the plot above (Fig.~\ref{Epeak_microvar}) is a further confirmation of a relation between the energy and the narrowness of the peaks.There is no clear division between short and long GRBs, but this could be due to the low statistics, especially for the short bursts. It will be possible to investigate further on this aspect when a larger number of such bursts will be detected. However, it must be noticed that the short bursts values seems to follow a power-law relation. The long GRBs too display shorter MVTs at higher peak energies, but the values are more spread out. Using a linear regression separately on the short and long GRBs, the two following relations are found: for the short GRBs the power-law goes as $\Delta t_{min,z} \propto E_{p,z}^{-0.85}$, for the long GRBs as $\Delta t_{min,z} \propto E_{p,z}^{-0.54}$. 

\subsubsection{The MVT in the Amati relation}

\begin{figure} [ht!]
\centering
\includegraphics[width=1.05\linewidth]{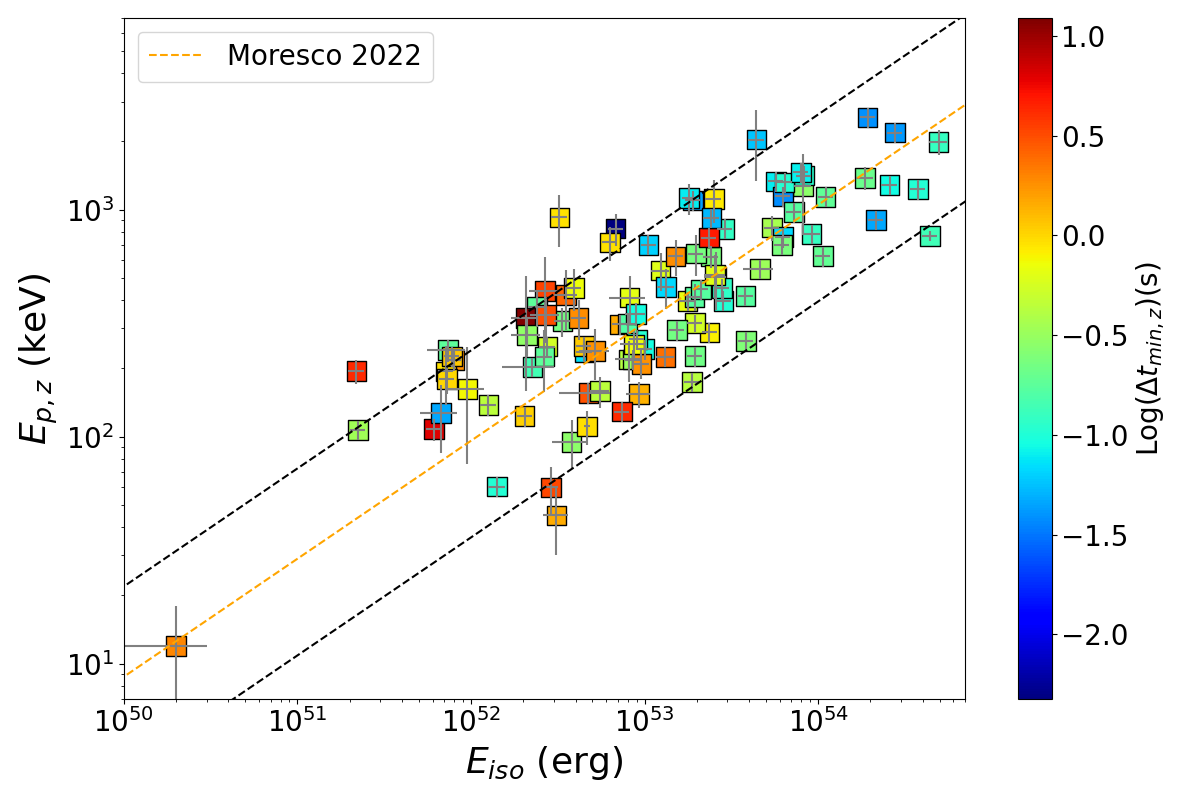}
\caption 
{ \label{Amati_microvar} 
Display of the Amati relation for 100 GRBs, with the corresponding MVT in the color scale. The relation found in \citet{2022LRR....25....6M}, on a similar but smaller sample than the Amati one, is reported with the dashed yellow and black lines.}
\end{figure}

The Amati relation \citep{2002A&A...390...81A} highlights the connection between the peak energy of a long GRB, in its cosmological rest frame, and its isotropic energy. In the previous sections, we have pointed out that the shortest variability of a GRB seems to be related to these two quantities; in particular, in both cases we observe a shorter MVT towards higher energy values. Exploiting the Amati relation, we can compare all these quantities together. To perform this step, we use the $E_{iso}$ and $E_{p,z}$ values provided directly by Lorenzo Amati (Amati et al, 2025, in preparation). This approach has two advantages: it enables us to include the GRBs whose spectra are not best-fitted with a Band function in the Fermi catalog, and we have the isotropic energy calculated over the $\SI{1}{\kilo\electronvolt}$--$\SI{10}{\mega\electronvolt}$ energy interval, so it is easier to compare with results from the literature where this range is typically employed. Of the 263 long GRBs selected from different satellites over 25 years, 100 are those for which we have calculated the MVT from the Fermi catalog, and which are reported in Fig.~\ref{Amati_microvar}.

Clearly, the relation individuated in the previous section is still present. Short MVT is found at high peak and isotropic energies, whereas at lower energies the MVT tends to be longer. An exception can be seen in the upper right part of the plot, for GRB160623 (red square). This exception can be explained by inspecting the Fermi GBM lightcurve of the GRB, which shows that it is a faint GRB. On the opposite side of the plot, GRB200826 (blue square) exhibits the exact opposite behavior. This GRB is peculiar since its $T_{90}$ would classify it as a short GRB, but it has been associated with a collapsar event for several other features \citep{2022ApJ...932....1R}. With the exception of these few cases, the trend seems to be well defined. 

\subsection{GRB total variability}
\label{subsec:variance}

An aspect that was not considered up to now is the variability of the burst over its entire duration. In order to compare with a recently performed study \citep{2024A&A...690A.261G}, we estimated the variability $V$ of the GRBs using the same formula, which was originally introduced by \citet{2001ApJ...552...57R}.

$$ V = \frac{\sum_{i=1}^n \left[\left(r_i-s_i\right)^2 - k\sigma_i^2\right]}{\sum_{i=1}^n \left(r_i^2-\sigma_i^2\right)}$$

where $r_i$ is the net count rate of the i-th bin of the raw GRB lightcurve, $s_i$ is the net count rate of the smoothed GRB lightcurve obtained through an Haar wavelet transformation, $\sigma_i$ is the Poisson noise associated to the i-th bin, and $k$ is a factor that corrects for the weight of the noise variance as described more in details in \citet{2024A&A...690A.261G}. In this case we used the entirety of the GRB lightcurve to calculate the smoothed version. The lightcurve binning was set at $64 (1+z)^{\beta}$ ms, in order to analyze every burst in its cosmological rest frame at the same binning time. The results shown in the following paragraph were obtained for $\beta = 0.6$.

\begin{figure} [ht!]
\centering
\includegraphics[width=\linewidth]{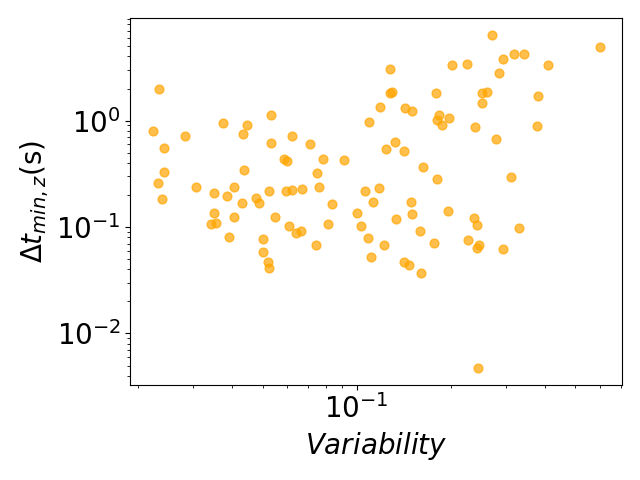}
\caption 
{ \label{microvar_var} 
Display of the relation between the MVT and the parameter V for long GRBs.}
\end{figure}

Being the main subject of this work, we compared the MVT with the newly calculated parameter $V$. We used only the long GRBs. To avoid including faint GRB that could bias the result, we only kept those with an overall SNR $> 30$. We display the result in Fig.~\ref{microvar_var} which shows no clear correlation between the MVT and $V$. We calculated the Pearson correlation factor, which confirmed the first impression, providing $r=0.27$ and a p-value of $3\cdot10^{-3}$.

\begin{figure} [ht!]
\centering
\includegraphics[width=\linewidth]{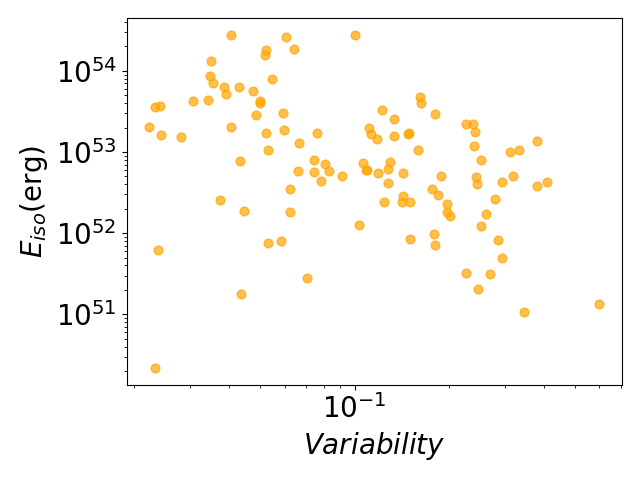}
\caption 
{ \label{Eiso_var} 
Display of the relation between the isotropic energy and the variability $V$ for long GRBs.}
\end{figure}

Finally, we checked if the variability $V$ defined above can be related to the isotropic energy (Fig.~\ref{Eiso_var}). We would expect no evident relation since we demonstrated in Sect.~\ref{subsubsec:Eiso} that the isotropic energy shows a certain degree of correlation with $\Delta t_{min, z}$. Nonetheless, a slight anti-correlation is found, with $r = -0.47$ (p-value $<10^{-3}$).

\subsection{New opportunities with future missions}
\label{subsec:microsec}

\begin{figure} [ht!]
\centering
\includegraphics[width=\linewidth]{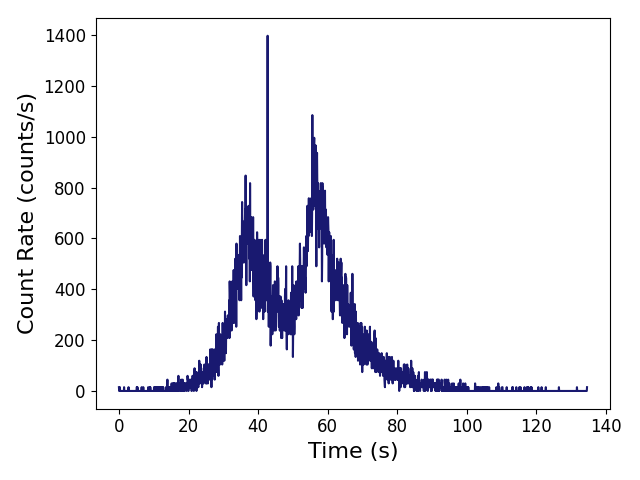}
\caption 
{ \label{simulated_GRB} 
Simulated GRB, containing a spike of microseconds duration.}
\end{figure}

From the results presented up to now, there is no proof of significative variability in the temporal structure of a GRB of the order of tenths of millisecond or lower. What if this is actually possible and we are limited by our instruments ? The GRB221009A \citep{2023ApJ...952L..42L}, with its extreme brightness and very narrow peaks, could be a possible candidate. However, because it saturated the detectors in the maximum activity phase, the analysis is rendered impossible since we are unable to observe the variation of collected photons by the detectors over time. This kind of ``monster" GRB nourishes the hope that it is possible to find a MVT of the order of microseconds. A new generation of mission is now being developed, and is hoped to be launched soon, to join and then replace the actual generation of mainly monolithic instruments, to create a constellation of collaborating satellites, like HERMES Pathfinder \citep{2021SPIE11444E..1RF}, GALI \citep{2021SPIE11444E..6ER}, GRBAlpha \citep{2021SPIE11444E..4VP}, while on the ground side CTA \citep{2013APh....43....3A} will provide an unprecedented view on the most energetic part of the GRB spectrum where we can expect to find the shortest variability.
For example, the HERMES Pathfinder nanosatellites will provide a better time resolution in the X-low $\gamma$ energy range compared to Fermi GBM or Swift, making it possible to explore the time structure of the lightcurve up to $\SI{1}{\micro\second}$ (maybe even lower). For all these reasons, it is therefore interesting to verify if the tool developed to obtain the MVT is capable of identifying these very low values. We perform this study on a simulated GRB, where we specifically added a pulse with a rise time of few microseconds in between two longer pulses, as shown in Fig.~\ref{simulated_GRB}. This could be the case of a very bright GRB, with a peak flux of $\sim \SI{100}{\photon\per\second\per\centi\meter^{2}}$, observed by a HERMES-like detector with a total effective area of few meters square (there should be about a few events per year of this kind observable by such a telescope).

\begin{figure*} [htb!]
\centering
\includegraphics[width=\columnwidth]{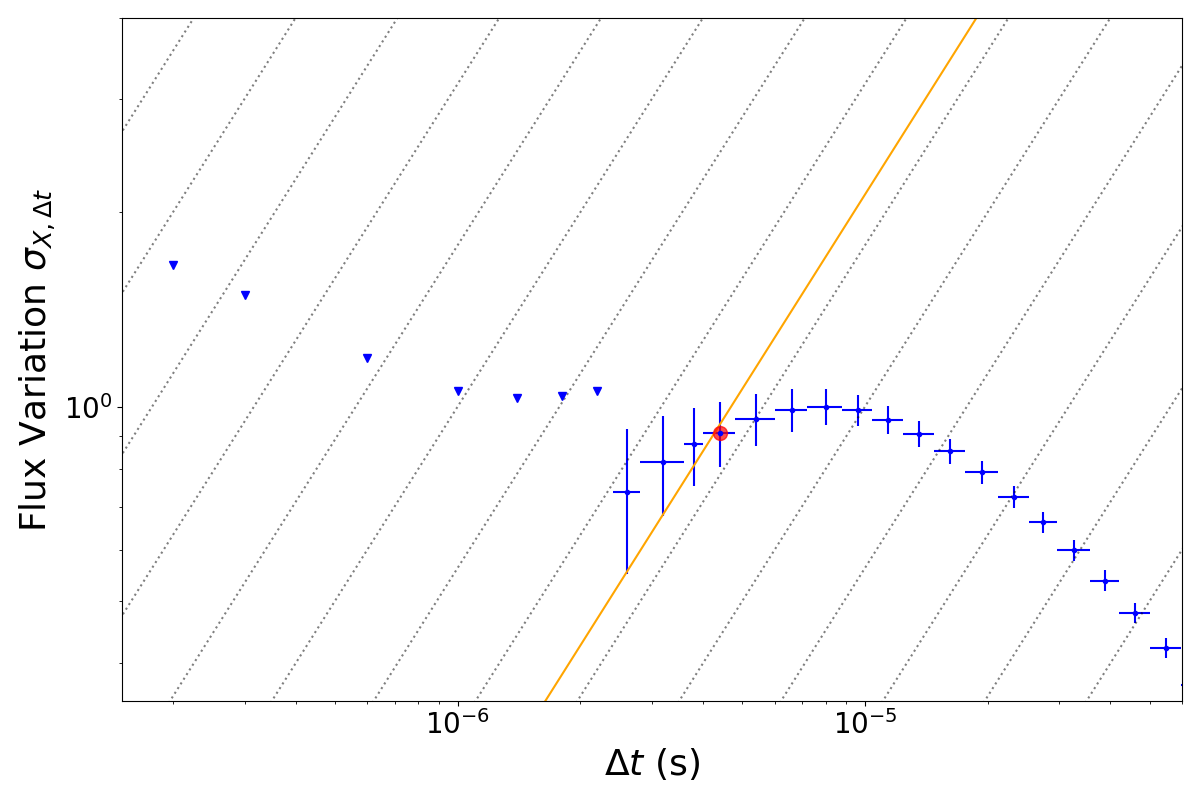}
\includegraphics[width=\columnwidth]{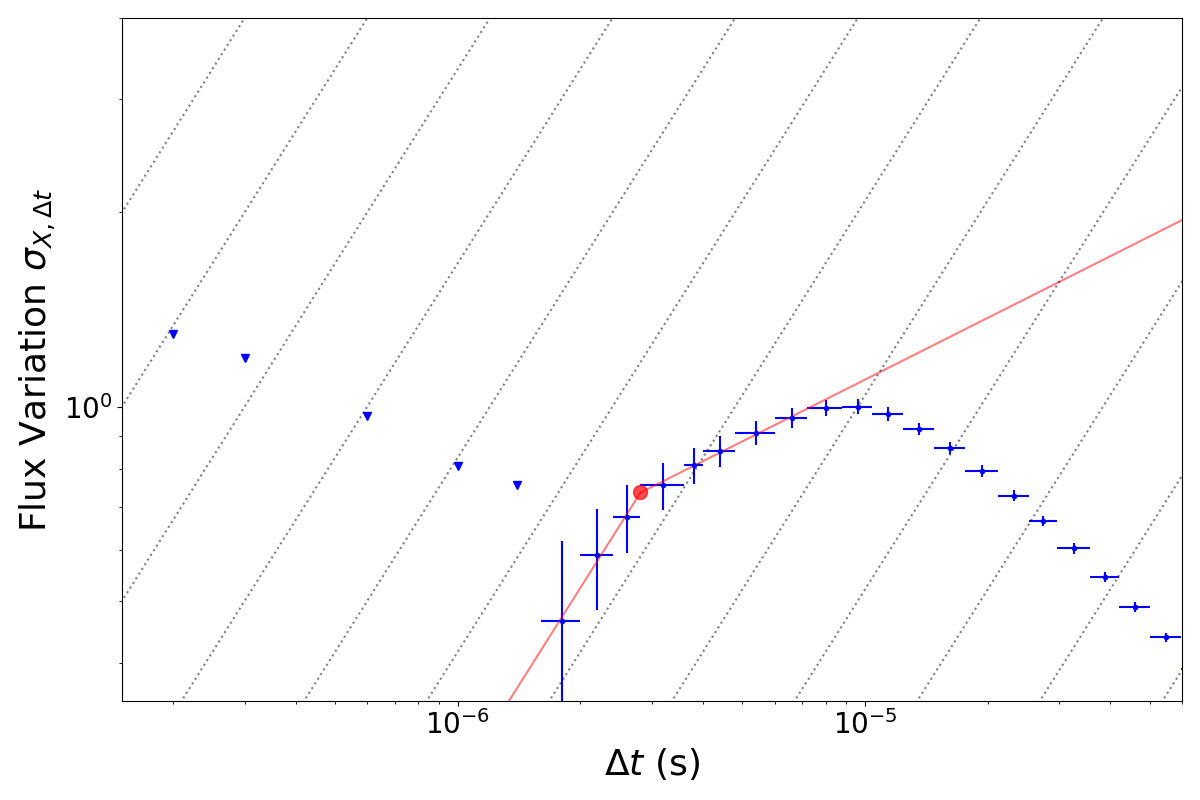}
\caption 
{ \label{Hermes_microvar} 
Scaleogram obtain with method 1 (left) and method 2 (right) for the simulated GRB. As in the majority of cases, the method 1 is more conservative.}
\end{figure*}

Performing the calculation of the shortest variability, we find a timescale between 3 and $\SI{5}{\micro\second}$, depending on the method used (see Fig.~\ref{Hermes_microvar}), as we were hoping in view of future observations.

\section{Discussion}

The concept of MVT of a gamma-ray burst has been thoroughly investigated in this work, using two different methods, one more conservative than the other. Applying these methods to a large set of GRBs ($> 3000$) observed by Fermi GBM, we found a median value for the MVT of about $\SI{0.74}{\second}$ ($\SI{0.58}{\second}$, with method 2) with a quartile deviation of $\SI{0.89}{\second}$ ($\SI{0.68}{\second}$) for long GRBs, and $\SI{78}{\milli\second}$ ($\SI{53}{\milli\second}$) with a quartile deviation of $\SI{65}{\milli\second}$ ($\SI{59}{\milli\second}$) for short ones, in the observer frame. A similar work was performed in \citet{2023ApJ...954L...5V}, using the Method 2 to evaluate the MVT. We compared the values obtained in that paper with those found in our work, finding them very similar or slightly lower, but still compatible within $3\sigma$. Furthermore, we can observe the similarities between Figure 2 of the mentioned paper and the right plot of Fig.~\ref{T90_MVT}. The shapes of the distribution of values are alike, even though the values of the MVT in the former are lower (as expected for what we said in Sect.~\ref{subsec:obs_frame}).

Moving to the cosmological rest frame (where we used only the method 1), for the ``redshifted" GRBs the MVT median values are shorter (as expected due to time dilation partially mitigated by other effects): $\SI{300}{\milli\second}$, with a quartile deviation of $\SI{410}{\milli\second}$, for long GRBs and $\SI{35}{\milli\second}$, with a quartile deviation of $\SI{24}{\milli\second}$, for short bursts. It is interesting to compare these values with a recent work, namely \citet{2023A&A...671A.112C}, that estimated the minimum time variability using a different method from wavelet. They define the minimum variability as the Full Width Half Maximum of the shortest statistically significant peak (established using an algorithm called MEPSA). Looking at the results obtained from the Insight-HXMT/HE data, that covers a similar energy range compared to Fermi GBM, we remark that they are similar, slightly higher for long GRBs. This comparison is encouraging to confirm the minimum variability values found in our work. Similar considerations can be made towards the \citet{2025A&A...702A..95M} work, which uses the same method of \citet{2023A&A...671A.112C} to evaluate the MVT over a sample of Fermi GBM GRBs similar to the one used for our work. The differences between the values found in this and our work are even larger: the FWHM method provides median MVT values larger by a factor of three for long GRBs and two for short ones, considering the MVT evaluated in the observer rest frame, as it is calculated in the work that we are comparing to.

Our research has initially demonstrated an energy-dependent trend in GRB lightcurve MVT: the MVT is shorter when a GRB is analyzed in an higher energy band as compared to the same GRB analyzed in a lower energy band. This is in accordance to previous studies that observed the sharpening of GRB lightcurves when using harder energy ranges, but this time it was performed in the cosmological rest frame. We also proved that accounting for the time dilation for the MVT, when moving to the cosmological rest frame, using a milder relation than simply $1+z$ gives consistent results within a $2\sigma$ confidence region. This was done considering a "Fenimore-like" correction, which is encouraged by the result reported just above. The slight negative slope displayed by the performed fit indicates the possible presence of other effects that might further reduce the time dilation impact on the MVT (e.g. tip-of-the-iceberg effect).

Subsequently, the anti-correlation between the MVT and the isotropic energy was demonstrated, at least for long GRBs, finding that higher isotropic energies correspond to shorter MVTs. Additionally we verified that this does not depend on the fluence of the transient event.

The peak energy study also provided an interesting relation with the MVT. In particular, for the short GRBs a linear regression was attempted obtaining the following relation: $\Delta t_{min,z} \propto E_{p,z}^{-0.85}$. For long GRBs the values are more scattered, but a similar analysis provides the relation $\Delta t_{min,z} \propto E_{p,z}^{-0.54}$. These relations should be further explored when a larger sample of GRBs with redshift will be available. The Amati relation allowed us to compare simultaneously all the three parameters ($\Delta t_{min,z}$, $E_{iso}$ and $E_{p,z}$) for every GRB. This further demonstrates the anti-correlations discussed in the previous paragraphs between the MVT and the isotropic energy, and between the MVT and the peak energy. In \citet{2025A&A...702A..95M} they report finding some weak anti-correlation for Type-II GRBs, with a power-law slope of $-0.19$. This is steeper than what we found for long GRBs (which largely correspond to the Type-II GRBs used). However, it must be noted that our MVT values are recalculated accounting for time dilation. If we keep the observer frame MVT value, the slope of the power law relation between MVT and $E_{p,z}$ becomes $-0.43$. In any case, both works agree in stating that the dispersion is very large. 

We also introduced the parameter $V$, taken from \citet{2001ApJ...552...57R} and \citet{2024A&A...690A.261G} to estimate the variability of the GRB, in particular the power of the variability on short timescales compared to the total power. We found no clear correlation between this parameter and the MVT. This is in accordance with what was found by \citet{2024A&A...690A.261G}, which concluded that there was no clear correlation between their MVT, calculated using the FWHM method (introduced previously in this section) and $V$. However, we observe a similar distribution of values of the MVT both for lower (few $10^{-2}$) and higher ($>10^{-1}$) values of $V$, contrarily to what was observed in the cited work. This might be due to the different way of estimating the smoothed version of the lightcurve.

From a modeling point of view, in \citet{1998MNRAS.296..275D}, the authors developed a model to reproduce as accurately as possible the GRB lightcurves. In this model, based on the internal shock scenario, they found that shells emitted with a random Lorentz factor on a time scale of milliseconds produce a realistic GRB lightcurve. This is exactly the lower limit of the timescales of minimum variability that we found in this work. This could be a proof of an internal shock scenario if we consider the MVT as a trace of the central engine activity. According to another, more recent, model, presented in \citet{2011ApJ...726...90Z}, where the GRB emission is magnetically dominated (Poynting flux), the MVT that we found could be related to another physical process, the presence of fast magnetic reconnection and turbulence. The variability timescale expected in the lightcurve in this scenario is between 1 and $\SI{100}{\milli\second}$, compatible with the results obtained in this work.

For GRBs with a massive star as progenitor, a detailed hydro-dynamical simulation was performed by \citet{2010ApJ...723..267M} to describe the jets propagation in the stellar progenitor and the surrounding circumstellar medium considering different conditions. Two different contributions to the total variability of the GRB lightcurve are highlighted in this work, the first one faster, due to the variability of the luminosity injected by the central engine in the jet origin, and the second one slower, due to the interaction of the jet with the stellar medium. The analysis reports also a strong correspondence between the variability of the central engine activity and the variability of the lightcurve following the first few seconds of activity of the simulated GRB. Combining our results with the conclusions provided by \citet{2010ApJ...723..267M}, considering a matter dominated model, we could affirm that the variability of the luminosity injected by the central engine can vary as fast as few milliseconds, if we consider the lowest values found. Average values of MVT would instead suggest a variability of the luminosity injection of the order of tens to hundreds of milliseconds.

Concerning the MVT-isotropic energy relation, its origin is not yet clear. It can be speculated that higher isotropic energies correspond to a more collimated beam. It has also been pointed out that the MVT is related to the activity of the inner engine. In an internal shock model, where two shells emitted at different times and with different velocities collide, the MVT can be used to estimate the jet angular opening distance, $L$, equal to $c \Delta t_{min,z}$. Considering the median value reported above, this corresponds to a value of $\SI{9e4}{\kilo\meter}$ for long GRBs and $\SI{1.6e4}{\kilo\meter}$ for short ones. This quantity is related to the distance of the central engine from the $\gamma$-ray production site, defined as $R \sim 2\Gamma^2L$ \citep{1996ApJ...473..998F}, where $\Gamma$ is the velocity of the shell resulting from the collision. Thus, we have $R \sim \SI{1.8e9}{\kilo\meter} \frac{\Gamma}{100}^2$ and $\SI{3.2e8}{\kilo\meter} \frac{\Gamma}{100}^2$ for long and short GRBs, respectively. Consequently the MVT is related to all of these physical parameters that regulate the inner engine.

As for the isotropic energy, also the possible relation of the peak energy with the MVT is unclear. It has been suggested in \citet{2016A&A...589A..97D} that a connection between the sharpness of the pulses and the peak energy of the GRBs could be due to magnetic reconnection events, which are stronger for higher energy events (i.e. higher peak energy), and could induce faster variability due to stronger turbulent regions, but this hypothesis would need deeper investigations.

What would we need to obtain more information ? First of all, as stated several times during this work, a larger sample of GRBs with redshift, especially for short GRBs, would provide a stronger statistics and more conclusive relations with the spectral parameters. Also, as explained in Sect.~\ref{subsec:microsec}, we currently are unable to properly explore the microsecond region for variability. When new instruments, with improved timing resolution, will provide new GRB data, it will be finally possible to rule in favor or against the existence of a microsecond variability, and consequently adapt the theoretical models. A topic that was not discussed in this work concerns the candidates long merger GRBs, i.e. GRBs displaying a duration longer than 2 seconds, but associated with a kilonova emission, which is the result of a compact merger event (e.g. GRB211211A). These GRBs further question the short/long division, as we have already discussed across this work. Unfortunately, the time interval of GRBs detection that was chosen at the beginning of this work excludes the two most prominent candidates. In a future work it would be very interesting to analyze them, and compare with other works results (e.g. \citet{2023ApJ...954L...5V}).

\section{Conclusions}

In conclusion of this work, we can affirm that:
\begin{itemize}
    \item the minimum observed variability in the cosmological rest frame is of the order of tens of milliseconds for short GRBs and hundreds of milliseconds for long GRBs, although it can go down as low as few milliseconds for some GRBs;
    \item the narrowness-hardness relation, already verified on the profile of the GRBs at larger timescales, is true also for the MVT, meaning that at higher energies we observe shorter minimum variabilities;
    \item the time dilation adapted according to the "Fenimore-like" correction was also demonstrated through the MVT of long GRBs with measured redshift, using a larger sample; short GRBs are still to few to be able to obtain a statistically significant result;
    \item we found evidence of an anti-correlation between the MVT and the isotropic energy, for long GRBs, while this is not observed for short ones; 
    \item similarly to the isotropic energy, also the peak energy of the GRBs seems to be anti-correlated to the minimum variability timescale, this time for both short and long GRBs (especially the formers); this would be in accordance with the Amati relation, that demonstrated that there is a correlation between the isotropic energy and the peak energy of a GRB;
    \item for long GRBs we confirmed the absence of correlation between the variability of the GRB, as defined here and in previously cited works, and the MVT, while finding a slight correlation with the isotropic energy;
    \item finally, in view of the future generation of satellites with better time resolution, this tool demonstrated to be able to identify short variabilities as low as few microseconds, if present.
    
\end{itemize}

\begin{acknowledgements}
    
This study was funded by the European Union Horizon 2020 Research and Innovation Framework Programme under the grant agreement HERMES-Scientific Pathfinder n. 821896 and from ASI-INAF Accordo Attuativo HERMES Technologic Pathfinder n. 2018-10-HH.0. The research leading to these results has also received funding from the European Union’s Horizon 2020 Programme under the AHEAD2020 project grant agreement n. 871158. We thank Dr. N. Butler for his availability to answer our questions and for providing information to calculate the variability of the GRBs. We thank Dr. L. Amati for providing us with a GRB dataset personally prepared, with all the spectral information. We thank the anonymous referee for comments and suggestions that improved the quality of our work.

\end{acknowledgements}

\bibliographystyle{aa}
\bibliography{bibliography.bib}

\begin{appendix}
\section{Selection effects verification}
\label{sec:appendixA}
We report in this appendix the analysis that was performed to verify the absence of selection effects in the relation found between the MVT and the isotropic energy. The first step consisted in evaluating the efficiency of our method to find the MVT, based on the peak flux of the GRB. To obtain this information we generated FRED pulses with peak flux ranging from $10^2$ to $10^5$ counts/second and with rising time going from $10^{-2}$ to $10^1$ seconds. The FRED pulses where generated according to the model reported in \citet{2005ApJ...627..324N}.

$$I(t) = A \lambda \exp(-\tau_1/t - t/\tau_2) \quad \textrm{for} \quad t>0$$

where $A$ is the peak rate, $\tau_1$ and $\tau_2$ are chosen so that $\tau_{peak} = (\tau_1 \tau_2)^{1/2}$ and $\lambda = \exp[2(\tau_1/\tau_2)^{1/2}]$. For each combination of peak flux and rising time (or MVT), a hundred FRED pulses where analyzed using the Method 1. The efficiency of detection of the expected MVT was then fitted using the same equation used in \citet{2023A&A...671A.112C} and \citet{2025A&A...702A..95M}, reported here:

$$ \epsilon = a \log_{10}\left(\frac{MVT}{s}\right) + b \log_{10}\left(\frac{PR}{counts/s}\right) + c$$

The values that we found for the three parameters are the following ones: $a = 1.62$, $b = 0.56$ and $c = 0.83$.

\begin{figure} [ht!]
\centering
\includegraphics[width=\linewidth]{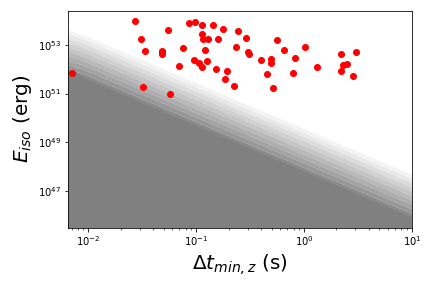}
\caption 
{ \label{detection} 
Example of the detection efficiency calculated through the simulation compared to the real GRBs observed by Fermi-GBM. This group of GRBs has redshift between 0.84 and 1.79. The shadowed region indicates the detection efficiency, starting from 0 (grey) to 1 (white)}
\end{figure} 

The second step consisted in creating new sets of GRBs starting from the values of the real GRB dataset with know redshift. Since the MVT-$E_{iso}$ relation was found only for the long GRBs, we proceeded with these. We followed the same procedure as in \citet{2023A&A...671A.112C} and \citet{2025A&A...702A..95M} since it seemed pretty satisfying for the purpose of this work. We divided logarithmically the GRBs in 9 groups based on their redshift. For each group we created "fake" GRBs in equal quantity to the real ones, assigning them the same isotropic energies of the real GRBs contained in that group, while the MVT was attributed performing a Gaussian kernel density estimation from the complete sample of MVTs of the long GRBs with know redshift. We also applied the same conditions to accept or reject the artificially created GRBs: from the equation estimated in the first step, we perform a Bernoulli test for each GRB, where the probability of success is given by the efficiency $\epsilon$; then we require that $E_{iso} / MVT < L_{max}$, where $L_{max}$ is the maximum luminosity observed in the real GRB dataset for each group.

\begin{figure} [ht!]
   \begin{center}
   \includegraphics[width=\linewidth]{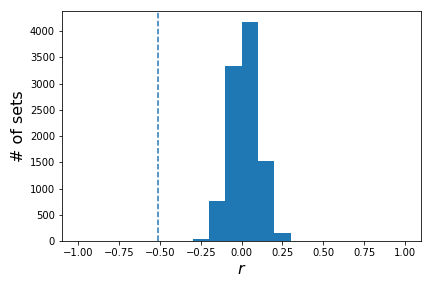}
   \end{center}
   \caption 
   { \label{corr_dist} 
    Distribution of the correlation values obtained for the $10^4$ artificial sets of GRBs. The blue dashed line correspond to Pearson correlation factor that was obtained with the real set of GRBs.}
\end{figure}

\vspace{5mm}
Once the full artificial set of GRBs is complete, we finally performed the Pearson correlation test between the MVTs and the isotropic energies. The final result is shown in Fig.~\ref{corr_dist}.

\end{appendix}

\end{document}